\documentclass[journal]{IEEEtran}
\usepackage{cite}
\usepackage{amsmath,amssymb,amsfonts}
\usepackage{algorithmic}
\usepackage{graphicx}
\usepackage{textcomp}
\usepackage{xcolor}
\usepackage{booktabs}

\def\BibTeX{{\rm B\kern-.05em{\sc i\kern-.025em b}\kern-.08em
    T\kern-.1667em\lower.7ex\hbox{E}\kern-.125emX}}
\begin{document}


\title{Shooting Neutrons at Neurons:\\ Radiation Testing of a Spiking Neural Network on Flash-Based FPGAs}

\author{Wim Nijsink, Bruno Endres Forlin, Amirreza Yousefzadeh, and Marco Ottavi\\
Dependable Computing Systems Group, CAES, University of Twente, The Netherlands\\
\{w.h.nijsink, b.endresforlin, a.yousefzadeh, m.ottavi\}@utwente.nl}

\maketitle

\begin{abstract}
Neuromorphic, or spiking, processors are increasingly being considered for use in harsh, radiation-prone environments such as space and avionics, where energy efficiency and graceful degradation are essential. In this study, we propose and experimentally validate a radiation-testing methodology specifically designed for neuromorphic processors that employ on-chip synaptic plasticity. We map the open-source ODIN SNN processor with Spike-Dependent Synaptic Plasticity (SDSP) onto the FPGA and expose it to a high-energy neutron beam while continuously monitoring MNIST classification accuracy and recording the synaptic state.

From these measurements, we extract SEU cross-sections for ODIN's synaptic memory and develop a calibrated fault model to inform a complementary fault-injection campaign. By comparing inference-only and online-learning configurations, we demonstrate that enabling SDSP can significantly extend the time to application-level failure and enable partial recovery from accumulated bit flips, with modest hardware overhead.
\end{abstract}

\begin{IEEEkeywords}
Radiation testing, flash FPGA, neuromorphic computing, spiking neural network, ODIN, SDSP, reliability, SEU.
\end{IEEEkeywords}

\section{Introduction}

Neuromorphic, or spiking, processors are increasingly being considered for energy-efficient computation in harsh and radiation-prone environments, such as satellites, planetary landers, and avionics platforms. Their event-driven operation, distributed state, and bio-inspired coding offer the advantage of graceful degradation under local faults, in contrast to traditional von Neumann architectures. However, in space and high-altitude applications, electronics are continuously exposed to ionizing radiation, which can induce both transient and permanent faults at the device level. To ensure that neuromorphic hardware can be successfully deployed in such missions, its robustness against radiation-induced soft errors must be assessed with the same rigor as that applied to conventional processors.

Radiation effects, such as Single Event Upsets (SEUs), Single Event Transients (SETs), and latch-up, are known to impact memory and logic at advanced technology nodes. Extensive test methodologies exist for microprocessors, FPGAs, and general-purpose accelerators \cite{forlin2025Ground, strojwas2019yield, bittel2024data}. Recent research has started to investigate the reliability and radiation tolerance of neuromorphic systems. This includes surveys of neuromorphic architectures designed for space applications \cite{naoukin2023survey}, proposed mitigation strategies and design guidelines \cite{schumann2022radiation, spyrou2025trustworthiness, picardo2022device, song2021dynamic}, as well as initial radiation studies conducted on proprietary platforms such as Intel Loihi \cite{scrofano2024radiation}. These efforts show that neuromorphic processors can demonstrate various behaviors under radiation, ranging from minor accuracy loss to complete failure. Therefore, it is crucial to implement mitigation strategies at the device, architectural, and algorithmic levels.

Despite advancements made, the current methods for assessing the robustness of neuromorphic systems remain inconsistent. Many research studies tend to focus on fault mechanisms at the device level without connecting them to metrics relevant to application performance. Additionally, some approaches rely solely on simulation-based fault injections using arbitrary error models that have not been validated against actual data \cite{schumann2022radiation, spyrou2025trustworthiness, picardo2022device}. Furthermore, neuromorphic systems that incorporate on-chip learning add extra complexity. While online adaptation can potentially compensate for some faults, it may also unintentionally amplify or accumulate other faults in ways that are not immediately obvious. Currently, there is no widely accepted, open methodology that (i) specifically targets neuromorphic processors with learning capabilities, (ii) operates on accessible hardware platforms, and (iii) establishes a reproducible connection between neutron-beam experiments and cross-layer fault-injection studies. This gap hinders the community's ability to compare different architectures, assess the advantages of learning-based fault mitigation, and design reliable neuromorphic systems for mission-critical applications.

In this study, we aim to address the existing gap by “shooting neutrons at neurons” and proposing a radiation testing methodology specifically designed for neuromorphic processors implemented on flash-based FPGAs. Flash-based devices maintain configuration integrity under irradiation, allowing us to isolate Single Event Upsets (SEUs) in the neuromorphic core rather than in the configuration memory. We apply this methodology to the open-source ODIN processor \cite{frenkel20180} with Spike-Dependent Synaptic Plasticity (SDSP) learning, which is mapped to a Microchip PolarFire SoC FPGA. The system is then exposed to a representative neutron spectrum at the ChipIR facility. Our setup includes online monitoring of MNIST classification, periodic dumping of synaptic memory for bit-flip analysis, and a calibrated SEU fault-injection campaign, with rates derived from the measured cross-section.

Our main contributions are as follows:
\begin{itemize}
    \item We created a radiation test framework for neuromorphic processors on flash FPGAs. This framework establishes a repeatable experimental process that integrates neutron-beam measurements with a fault model and tools adapted from advanced soft-core test methodologies \cite{forlin2025Ground}, specifically tailored for neuromorphic workloads and memory systems.

    \item This study presents the first systematic examination of radiation effects on an open neuromorphic core implemented on a flash FPGA. Utilizing the ODIN tool with SDSP on a PolarFire® SoC, we evaluate how single-event upsets (SEUs) in the synaptic and neuronal states impact classification accuracy. We measure both the immediate and cumulative effects of these disruptions under realistic neutron flux conditions.

    \item We performed a quantitative assessment of learning-enabled robustness. By comparing inference-only configurations with online-learning configurations, we discovered that enabling on-device learning could potentially replace the expensive Triple Modular Redundancy (TMR) in synaptic memories.

    \item A scalable and open approach to designing reliability-aware neuromorphic systems is proposed. This methodology serves as a framework for future research on larger neuromorphic systems, allowing for calibrated fault-injection experiments and a comparative assessment of mitigation techniques. It aims to connect the fields of hardware dependability and neuromorphic computing.
\end{itemize}

\section{Test Methodology}

The proposed framework bases its experimental setup and methodology on \cite{forlin2025Ground}. The methodology defines both a \textbf{fault model} and an \textbf{experimental setup} for reproducible radiation testing of neuromorphic systems.

\subsection{Fault model and hardware platform}

The system is susceptible to the typical spectrum of radiation-induced effects, including single-event upsets (SEUs), single-event transients (SETs), and single-event latch-ups (SELs). Our analysis focuses on SEUs, as SETs are difficult to isolate in practice and ultimately manifest similarly on stateful devices as single- or multi-bit memory corruptions. SELs were not observed during this campaign.

The experimental setup therefore, consists of a flash-based FPGA running the ODIN core under continuous neutron irradiation at ChipIR, with protected control logic, UART-based stimulus and response streaming, and full internal state observability.

\begin{figure*}
    \centering
    \includegraphics[width=0.9\linewidth]{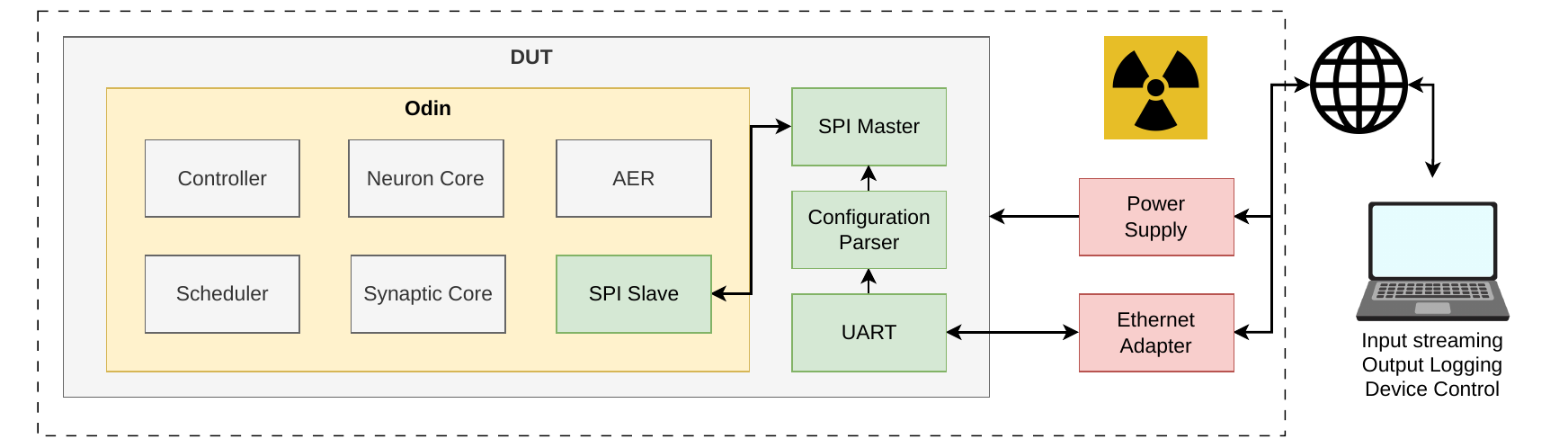}
    \caption{Experimental setup. Modules in green have automatic TMR applied; red modules are part of the test fixture.}
    \label{fig:fixture}
\end{figure*}

These fundamental faults can lead to observable system-level behavior such as accuracy degradation, hangs, resets, and crashes. Since evaluating control logic robustness is not the goal of this experiment, the control modules (UART, SPI master, configuration parser) are protected using triple modular redundancy (TMR) to prevent loss of communication during irradiation, as shown in Fig.\ref{fig:fixture}. The ODIN core is implemented in Verilog and synthesized using Microchip’s Libero SoC toolchain. Additional modules provide:

\begin{itemize}
    \item Real-time configuration and reset through a UART interface.
    \item Synapse and neuron memory read/write access for full state observability.
    \item Integration with the RadHelper\footnote{https://github.com/radhelper/radhelper-embedded} host tool for streaming inputs and collecting results.
\end{itemize}

\begin{figure}
    \centering
    \includegraphics[width=0.9\columnwidth]{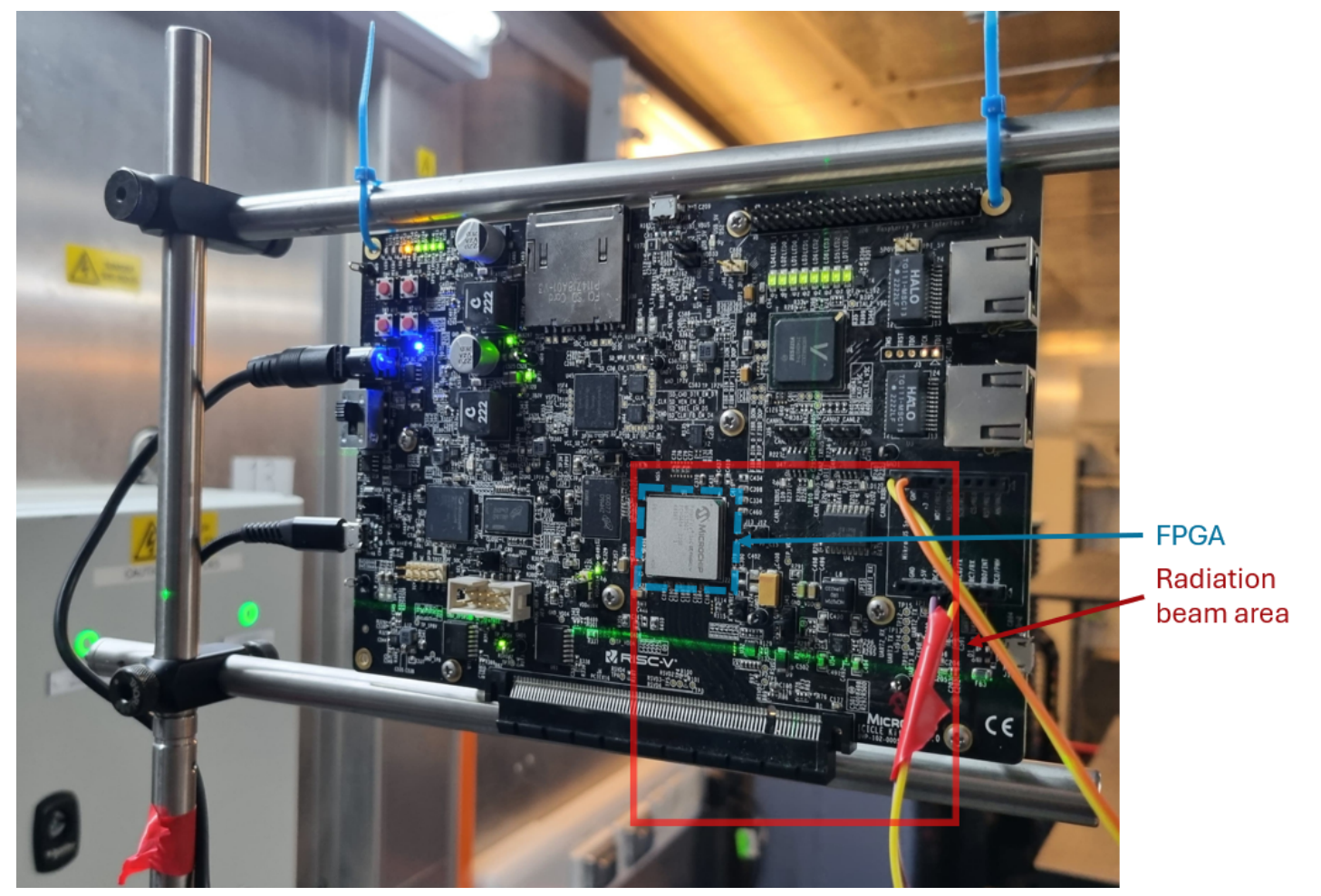}
    \caption{Experimental setup for radiation testing at ChipIR.}
    \label{fig:setup}
\end{figure}

The FPGA is programmed once and remains operational for the entire experiment without reconfiguration, leveraging the inherent radiation tolerance of flash-based devices. The beam setup is shown in Fig.\ref{fig:setup}. Input stimuli are streamed through the UART interface, and outputs are read back directly.

At the time of testing, the ChipIR beamline permitted a 10\,cm~$\times$~10\,cm beam window, highlighted in red in Fig.\ref{fig:setup}. The board was positioned to avoid most critical components while keeping the FPGA within the homogeneous neutron field.

\subsection{Test execution and metrics}

During beam exposure, we stream preprocessed MNIST images (16$\times$16 pixels, spike-encoded) to the FPGA. We define a single run as one full pass over a fixed image set (either 6k or 60k images), followed by logging and, optionally, a memory dump. A 6k-image run takes approximately 20–22 minutes end-to-end (UART streaming, inference, and logging). To accumulate sufficient fluence for statistics, we repeated the 6k run back-to-back, resulting in 21 full 6k runs per configuration, i.e., 7h and 10min total beam-on execution per 6k configuration. Similarly, a 60k-image run takes approximately 4 hours, and we repeated it 6 times for a total of 27h and 37min.

In total, each short configuration was executed for 7 hours and 10 minutes (14 hours combined), corresponding to 21 full executions of 6k images per configuration. The long configuration was executed for 27 hours and 37 minutes, corresponding to 6 full executions of 60k images.

The facility beam logs provide the neutron flux $\Phi$ inside the irradiation chamber. The total fluence $\Psi$ was computed from the measured flux and the cumulative device active time for each configuration, as summarized in Table~\ref{tab:fluence_runtime}.

\begin{table}[ht]
    \centering
    \caption{Radiation fluence and runtime for each experimental configuration.}
    \label{tab:fluence_runtime}
    \begin{tabular}{lcc}
        \toprule
        Configuration & Runtime (h:m) & Fluence ($\Psi$ [particles/cm$^{2}$]) \\
        \midrule
        6k Learning-enabled       & 7:10  & $1.14\times10^{11}$ \\
        6k inference-only   & 7:10  & $1.18\times10^{11}$ \\
        60k Learning      & 27:37 & $4.03\times10^{11}$ \\
        \bottomrule
    \end{tabular}
\end{table}

From the number of observed bit-flips in the synaptic memory, the SEU cross-section $\sigma$ is calculated as
\begin{equation}
\sigma = \frac{N_{\text{errors}}}{\Psi}.
\end{equation}

\section{Results}

\subsection{In-beam behavior (ChipIR) and short/long runs}

\begin{figure}
    \centering
    \includegraphics[width=1\linewidth]{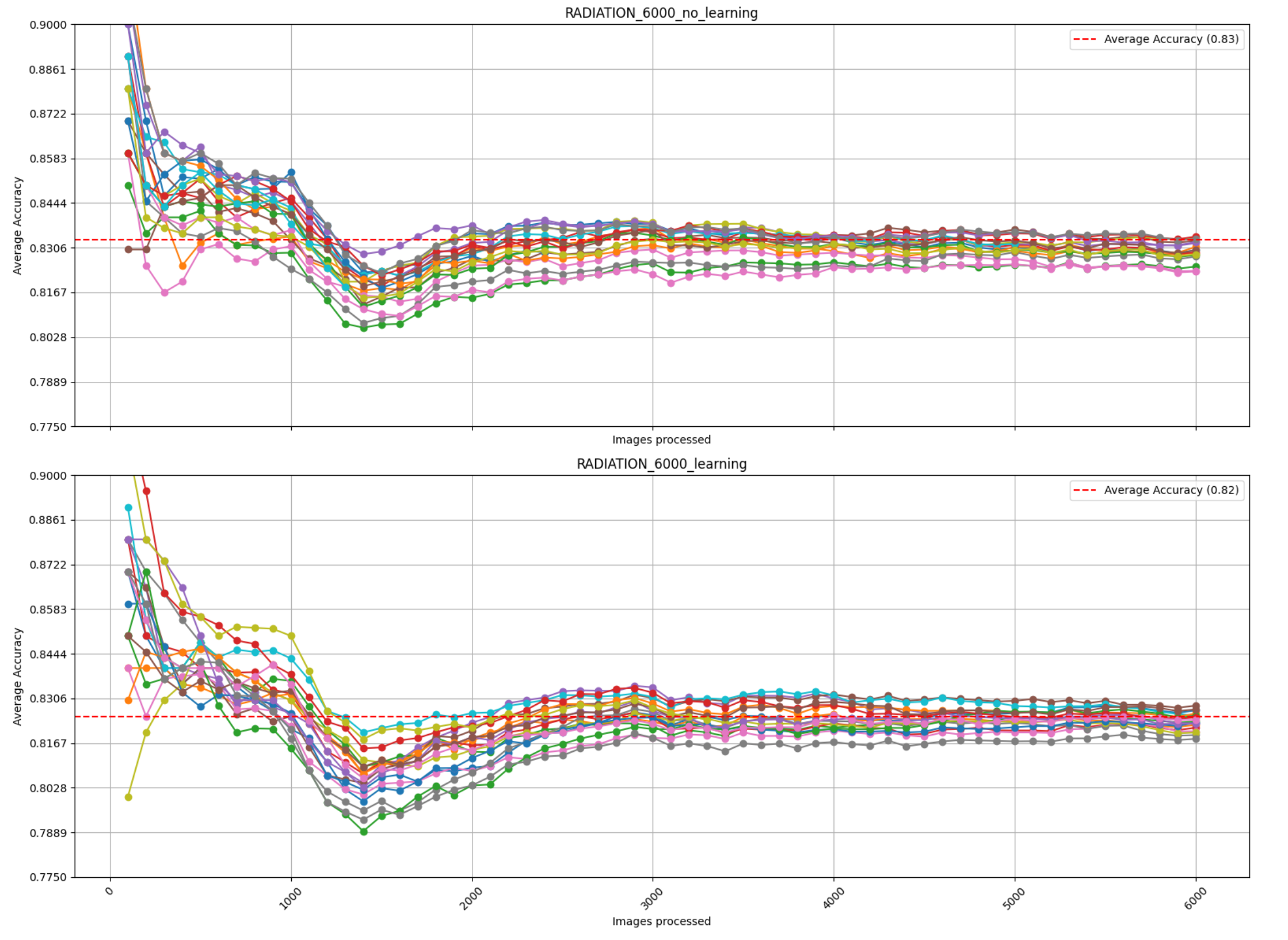}
    \caption{Accuracy under radiation over time for 6000 images. Top: no learning, bottom: with learning}
    \label{fig:5.14}
\end{figure}

\begin{figure}
    \centering
    \includegraphics[width=1\linewidth]{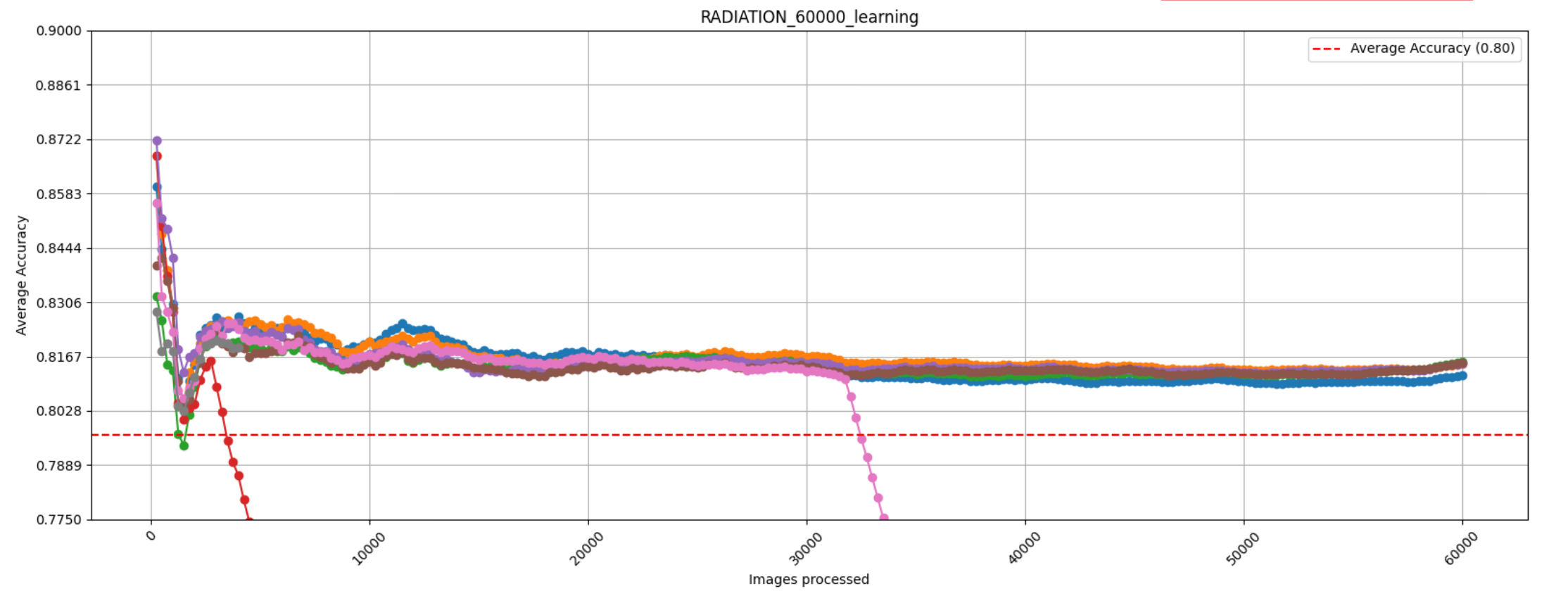}
    \caption{Accuracy under radiation for 60000 with learning}
    \label{fig:5.15}
\end{figure}

During neutron-beam exposure at ChipIR, ODIN maintained high initial MNIST accuracy ($\sim$82\%) in short in-beam runs. For the 6k-image experiments ($\sim$20–22 minutes per run), the in-beam accuracy remained stable over time for both configurations, as shown by the accuracy traces in Fig.\ref{fig:5.14}. 

For the 60k-image in-beam experiments ($\sim$4 hours per run), only the online-learning configuration was executed under the beam due to limited irradiation time. The corresponding accuracy-over-time traces are shown in Fig.\ref{fig:5.15}. Two of the six in-beam runs exhibited rapid, large accuracy degradation.

Overall, short exposures show small deviations and stable accuracy (Fig.\ref{fig:5.14}), while longer exposures occasionally produce abrupt drops (Fig.\ref{fig:5.15}). Such events may involve faults beyond the synaptic memory upsets used in fault-injection simulations.

\subsection{Bit-flip probability in the FPGA synaptic memory (cross-section)}

To quantify radiation-induced upsets, synaptic-memory dumps were compared before/after irradiation to count bit changes attributable to the beam. 
To avoid mistaking changes caused due to the learning of new weights, we evaluate the device's cross-section based on the inference-only configuration. The beam logging analysis (fluence integration) yields a corrected fluence $\Psi_{\text{corr}} = 1.18\times 10^{11}$ particles/cm$^{2}$ using the inference-only 6k-image in-beam reference, indicating that all observed changes are radiation-induced. This results in a total synaptic-memory cross-section for this ODIN configuration of:

\begin{equation}
\sigma = \frac{492}{1.18\times 10^{11}} = 4.17\times 10^{-9}\ \text{cm}^2. 
\end{equation}

The corresponding mean time to a synaptic-memory error under ChipIR conditions is $\approx 46s$.
Because ODIN uses only a subset of the full synaptic array in this configuration, we compute an “active/relevant memory” factor (10 enabled neurons): 10,240 relevant bits out of 262,144 total bits ($\eta \approx$ 3.9\%), which leads to an effective average single-event rate of $\approx$ 1185s ($\sim$20 min), i.e., roughly 3 relevant bit-flips per hour at ChipIR.    

\subsection{Accuracy degradation over time on the 10k-image test set (with vs. without learning)}

Since long in-beam testing at scale is time-constrained, we use fault-injection simulation calibrated to the measured ChipIR upset rate to evaluate long-duration behavior. Time is discretized into “periods” where each period corresponds to an \textit{equivalent} beam-exposure duration; after each period, ODIN is evaluated on the full 10k-image MNIST test set. A full round of unsupervised on-device learning (epoch) is executed on ODIN once after each period to measure the effect of learning on accuracy. 

\subsubsection{Shorter simulated periods (20–40h)} 

\begin{figure}
    \centering
    \includegraphics[width=1\linewidth]{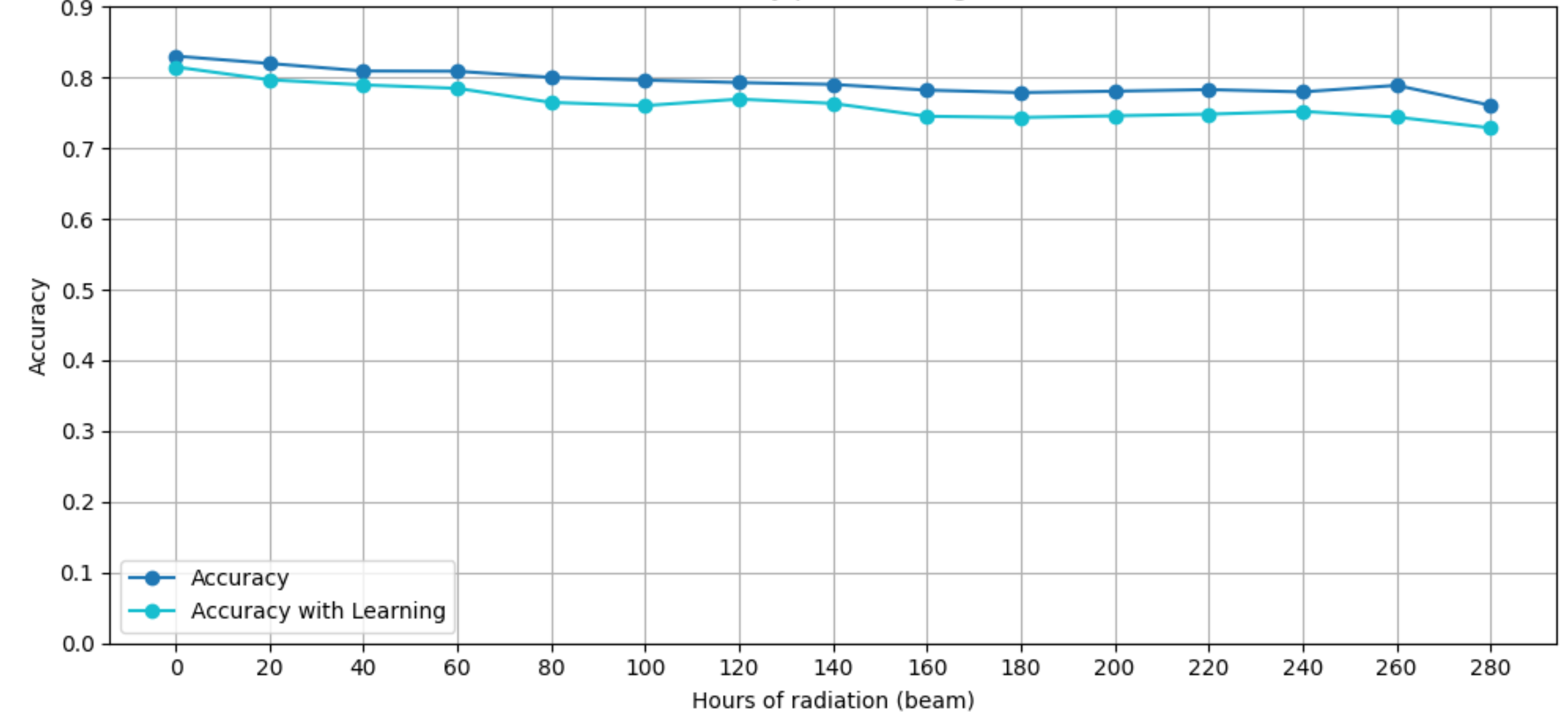}
    \caption{Accuracy over time for 10000 images with simulated radiation. One full learning epoch happens every 20 hours.}
    \label{fig:5.17}
\end{figure}

\begin{figure}
    \centering
    \includegraphics[width=1\linewidth]{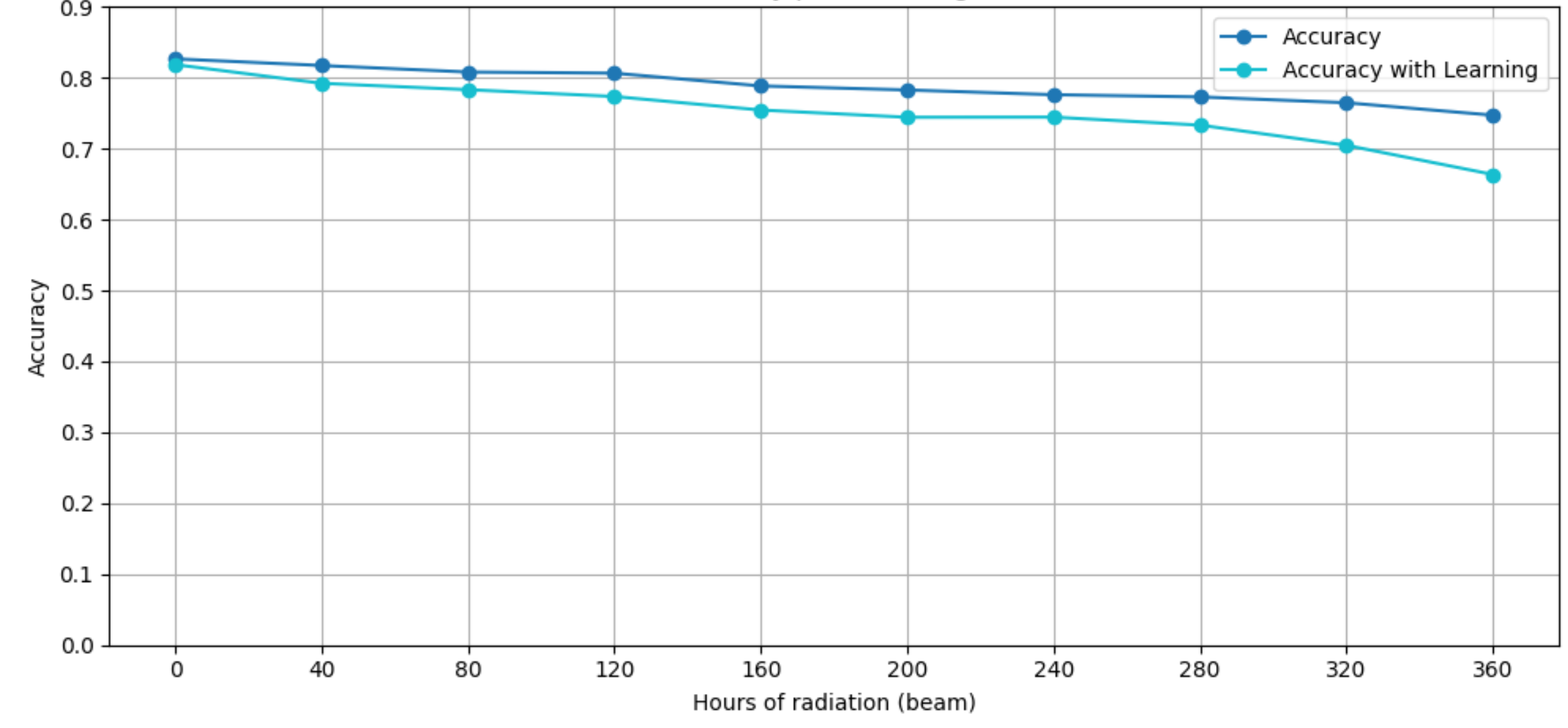}
    \caption{Accuracy over time for 10000 images with simulated radiation. One full learning epoch happens every 40 hours.}
    \label{fig:5.18}
\end{figure}

With 20h periods, learning-enabled and inference-only remain close (Fig.\ref{fig:5.17}). With 40h periods, the accuracy drop becomes more pronounced due to longer exposure between fault injections (Fig.\ref{fig:5.18}).

\subsubsection{Longer simulated periods (120h)} 

\begin{figure}
    \centering
    \includegraphics[width=1\linewidth]{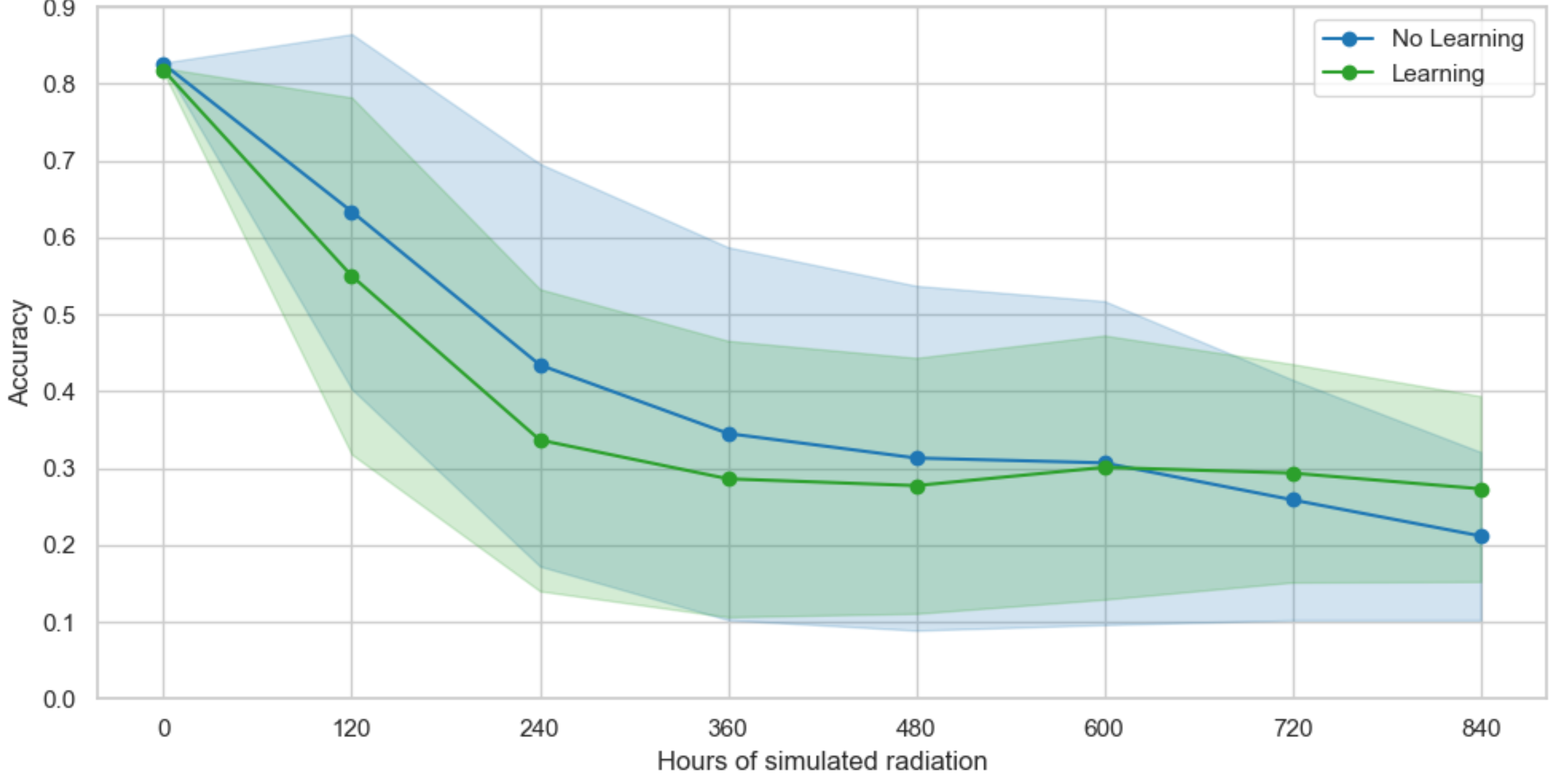}
    \caption{Accuracy over time for 10000 images with simulated radiation. One full learning epoch happens every 120 hours. To quantify variations, we run 45 independent simulations.}
    \label{fig:5.26}
\end{figure}

\begin{figure}
    \centering
    \includegraphics[width=0.75\linewidth]{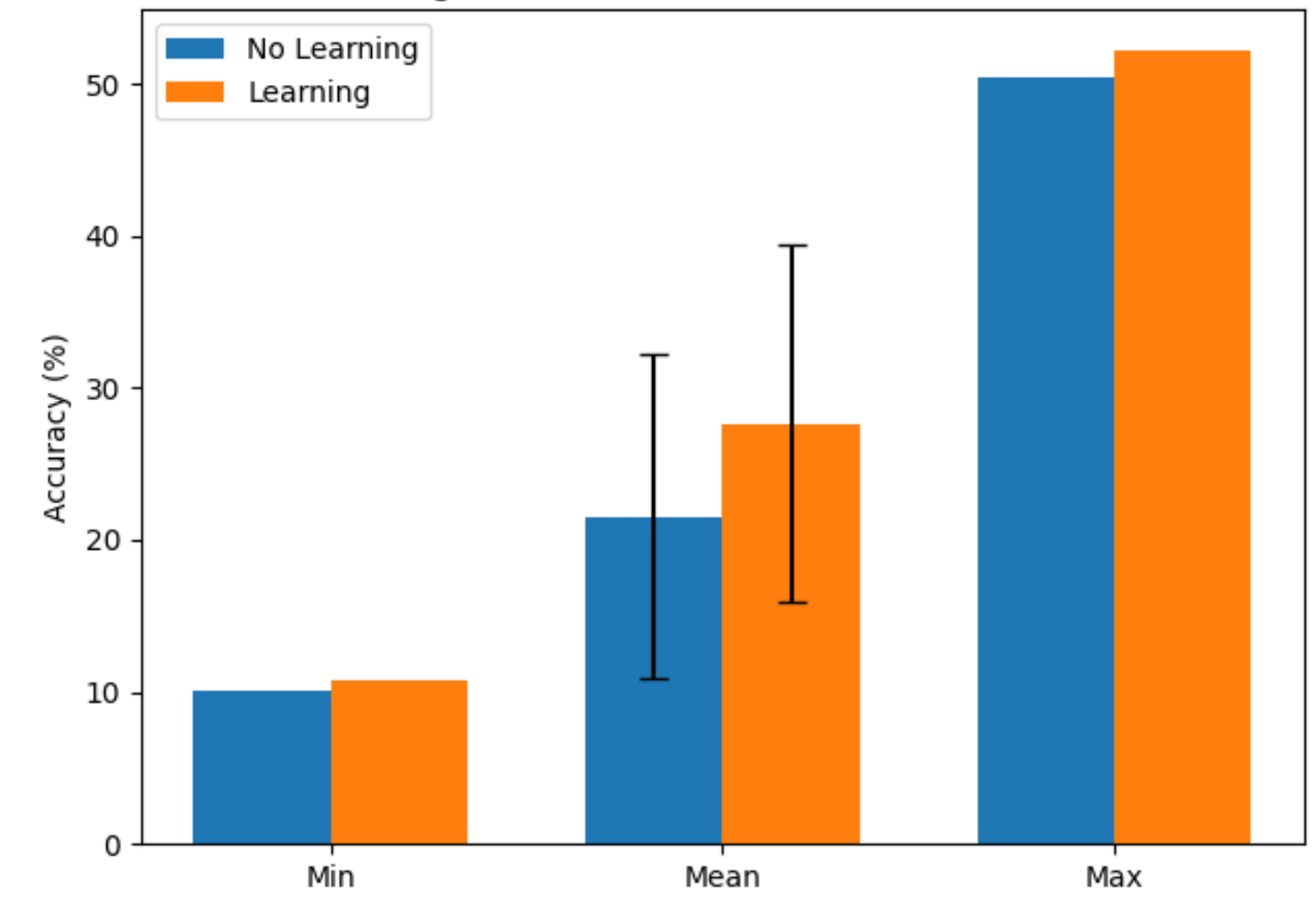}
    \caption{Min, mean, and max accuracy on 10000 images with simulated radiation after 840 hours}
    \label{fig:5.10}
\end{figure}

With 120h periods, both configurations can suffer early drops; however, across runs, learning tends to stabilize later and can maintain a higher accuracy than the inference-only configuration after multiple periods (Fig.\ref{fig:5.26}). The end-of-run accuracy statistics on the 10k-image test set are given in Fig.\ref{fig:5.10}. 

\subsection{Hardware resource comparison}


Table~\ref{tab:resources} summarizes the FPGA implementation cost of three ODIN configurations on PolarFire SoC. Enabling SDSP online learning increases the compute/control logic in the synaptic update path, raising LUT usage from 62k to 75k (+21\%) and FFs from 53k to 59k (+11\%), while leaving the embedded SRAM footprint unchanged at 36 KB. This indicates that SDSP primarily adds update/control circuitry and datapath logic rather than additional on-chip memory.
In contrast, applying TMR to the neuron and synapse memories has only a modest impact on logic—LUTs increase from 62k to 65k (+4–5\%) and FFs remain essentially unchanged (53k), because the added voters/control are small relative to the overall design. The dominant overhead is in memory capacity: the embedded SRAM increases from 36 KB to 108 KB (a 3$\times$ increase), consistent with triplicating the protected memory arrays.

In neuromorphic chips, the area is typically dominated by memory \cite{memory_wall_neuromorphic}. SDSP provides fault tolerance by continuously refreshing and re-adapting synaptic states via activity-driven updates, which can partially compensate for accumulated bit flips over time at a comparatively small memory cost. TMR remains valuable when instantaneous correction of the critical state is required, but its cost scales linearly with memory size; for large synaptic arrays, memory triplication can quickly become the dominant constraint compared to the modest logic overhead of learning.

\begin{table}[t]
\centering
\caption{Hardware overhead comparison (PolarFire SoC FPGA).}
\label{tab:resources}
\begin{tabular}{lcccc}
\toprule
Configuration & LUTs & FFs & SRAM \\
\midrule
ODIN baseline & 62k & 53k & 36KB \\
ODIN + SDSP & 75k & 59k & 36KB \\ 
ODIN + TMR & 65k & 53k & 108KB \\
\bottomrule
\end{tabular}
\end{table}

\section{Discussion and Future Work}




The experiments show that flash-based FPGAs are a reliable and reproducible platform for neuromorphic radiation studies. Their lack of configuration memory upsets allows researchers to focus on the corruption of application-relevant states, such as synaptic and neuronal states, instead of FPGA reconfiguration failures. This characteristic is crucial for cross-layer analysis: the device can operate for long, uninterrupted periods, the communication infrastructure can be independently protected, and the internal state can be periodically monitored and correlated with application-level metrics.

The stability observed in short exposures suggests that a moderate number of single-event upsets (SEUs) during exposure do not immediately result in functional failure. This supports the idea that neuromorphic representations can deteriorate gradually. In contrast, the infrequent but severe accuracy drops observed in long tests indicate that some failure modes likely go beyond the synaptic-memory SEU mechanism modeled in the calibrated fault-injection campaign. This observation highlights the need to enhance observability and expand the fault model to include additional architectural states, ensuring that simulated campaigns reflect not only gradual accuracy degradation but also sudden "cliff-edge" events.

A key finding of this work is the quantified advantage of enabling Spike-Dependent Synaptic Plasticity (SDSP) compared to operations that rely solely on inference. This comes with only a modest increase in logic overhead. In experiments with long-duration calibrated injections, online learning tends to slow down the decline in accuracy and can partially recover from accumulated state perturbations. One way to interpret this is that SDSP provides a controlled method of state refresh: synaptic values are continuously adjusted based on ongoing activity. This adjustment can help combat some persistent bit-flip corruptions, particularly when these corruptions push weights toward less effective configurations. 

Importantly, this advantage is achieved without the significant replication costs associated with full memory Triple Modular Redundancy (TMR), showing that algorithmic plasticity can serve as a low-overhead resilience mechanism. However, it’s essential to note that SDSP is not a universal solution. Its effectiveness depends on the operating conditions, the frequency of updates, and whether faults occur in areas where learning can "re-train" around them. Therefore, learning should be viewed as a complementary mitigation layer that reduces time-to-failure across many fault trajectories, rather than as a means of completely eliminating radiation sensitivity.

The methodology presented clarifies how to connect beam tests with fault-injection studies in a reproducible manner. By extracting a synaptic-memory cross-section under a controlled configuration and using it to calibrate injection rates, this study offers a principled alternative to the arbitrary error-rate assumptions commonly found in simulation-only robustness evaluations. This cross-layer connection—beam → upset statistics → calibrated injection → application-level robustness—facilitates fair comparisons across different architectures and mitigation techniques, as long as the modeled state elements correspond to the primary failure modes observed during beam testing.

Limitations and potential threats to validity should be acknowledged. First, the calibrated model is based on observed synaptic memory bit flips within a specific configuration and workload. The upset rate and impact distribution may vary when (i) more neurons and synapses are activated, (ii) activity patterns change, or (iii) additional architectural states become the primary contributors. Second, the difference between the gradual degradation seen in injection studies and the occasional abrupt drops observed in-beam suggests that the current model does not fully account for all relevant error modes. Third, while the MNIST dataset serves as a convenient and repeatable benchmark, mission workloads may demonstrate different sensitivities to state perturbations (for example, in temporal tasks, sensor fusion, and control systems). These limitations do not undermine the overall conclusions but instead define the scope of the findings: the current results most strongly support (a) the feasibility of using flash-FPGA neuromorphic systems under irradiation with sustained observability and (b) the potential robustness of plasticity as a low-overhead mitigation strategy.

Future developments will extend this framework to larger neuromorphic systems and explore:
\begin{itemize}
    \item Fine-grained in-beam logging of neuron and synapse upsets.
    \item Quantization-aware training to improve post-radiation accuracy.
    \item Automation for parallel multi-FPGA test campaigns.
\end{itemize}

\section*{Acknowledgment}

The authors acknowledge the ChipIR team at Rutherford Appleton Laboratory for their support during beam tests.

\bibliographystyle{IEEEtran}
\bibliography{references}

\end{document}